# LETTER TO THE EDITOR
# NAA & JFK: CAN REVISIONISM TAKE US HOME?

By John E. Fiorentino

Occasionally during the course of the human learning experience we are faced with an anomaly. An aberration of sorts, which try as we might, defies appropriate classification. The recent paper by Spiegelman et al.—*Chemical and forensic analysis of JFK assassination bullet lots: Is a second shooter possible?*—is one such aberration. It is riddled with both misconceptions and errors of fact.

Purporting to cast doubt on the NAA (neutron activation analysis) work conducted by Dr. Vincent Guinn in the investigation of the assassination of President John F. Kennedy, it fails miserably.

The paper offers two central conclusions, one which is demonstrably false, and the other which is specious.

The authors opine; "*If the assassination fragments are derived from three or more separate bullets, then a second assassin is likely, as the additional bullet would not be attributable to the main suspect, Mr. Oswald.*"

This statement relating to the likelihood of a second assassin based on the premise of three or more separate bullets is demonstrably false. The available evidence indicates that Oswald fired three shots, one of which is believed to have missed. However, on the off chance that all three shots hit (even though there is absolutely no other supporting forensic evidence for such a notion) those three shots alone in no way would indicate then that "a second assassin is likely." The authors' erroneous conclusion was achieved because they have either been misled (which I personally believe is the case) or they simply aren't familiar with the evidence.

The second fatal flaw is the use of a rather uncomplicated formula based on Bayes Theorem. Through this formula the authors conclude: "thus the critical ratio = .53 over .80. *Since this ratio is less than 1, Dr. Guinn's testimony that the evidence supports two and only two bullets making up the five JFK fragments is fundamentally flawed.*"

Several problems immediately present themselves. The authors are using the method known as Bayes Factors; this is legitimate and the equation they quote is derived from Bayes' law. The first term is the posterior odds of 2 bullets versus 3; the second term is what is known as the "likelihood ratio," not "ratio of probabilities" as the authors claim. It makes no sense to talk about probabilities of evidence; but this is a matter of fundamentals not methodology. The third term is the prior odds. Intriguingly, the prior odds the authors mention pertain to multiple shooters versus a single shooter. It should be the prior odds of 2 bullets versus 3 or more, because the posterior odds pertain to these events. However, nowhere in the paper do you see what prior odds were used. The ratio 53 to 80 is, from a Bayesian point of view effective only when it is modulated by the prior odds.

By using a misguided and biased approach to this very important evidence, Spiegelman et al. have effectively negated their findings.

In a nutshell the ONLY evidence of any other "bullets" MUST come from the several minute particles of lead recovered from the victims and the crime scene, as we have evidence of 2 bullets which are ballistically matched to the Oswald rifle to the exclusion of all other weapons in the world. So, IF there is another bullet the ONLY evidence of its passing is a particle (or particles) of lead. Unfortunately, for that idea, this "bullet" must have hit someone or something in the limousine. Unfortunate again is the fact there are no wounds in either victim to which it could be attributed, nor any damage observed in the limousine, to which it could be attributed. So, essentially we are left with a "bullet" which didn't hit either victim, didn't hit anywhere in the limousine, and only left behind a particle (or particles) of lead as evidence of its passing. Now, I'm not exactly sure of the "statistical probability" for that event, but my feeling is the answer lies somewhere outside the domain of either statistics, or chemical analyses. Perhaps it might come from the examination of tea leaves, lines on the palm of one's hand, or bumps on the head. But wherever you may find it, it won't be "science."

Fiorentino Research
PO Box 324
Oakhurst, New Jersey 07755
USA
E-mail: johnfiorentino@optonline.net
URL: www.fiorentinoresearch.com